\documentclass{article}

\usepackage{amssymb,amsfonts,amsmath,stmaryrd}
\usepackage{empheq}
\usepackage{cite,enumerate,float}
\usepackage{color}
\usepackage{tikz}
\usetikzlibrary{arrows,snakes,backgrounds}

\def\be{\begin{eqnarray}}
\def\ee{\end{eqnarray}}

\definecolor{red}{rgb}{1,0,0}
\definecolor{orange}{rgb}{1,0.5,0}
\definecolor{violet}{rgb}{0.7,0,1}

\hoffset= - 1.0in         

\newcommand\br[1]{
  \left(#1\right)
}
\newcommand\hpsi[0]{ \hat{\psi} }

\newcommand\ulambda[0]{
  {\underline{\lambda}}
}
\newcommand\umu[0]{
  {\underline{\mu}}
}
\newcommand\uDelta[0]{
  {\underline{\Delta}}
}
\newcommand\uz[0]{
  {\underline{z}}
}
\newcommand\ux[0]{
  {\underline{x}}
}

\newcommand\lsb[0]{\left [}
\newcommand\rsb[0]{\right ]}
\newcommand\sbr[1]{\lsb #1 \rsb}
\newcommand\tpsi[0]{\tilde{\psi}}
\newcommand\tff[0]{\text{ff}}

\newcommand\ssbr[1]{\llbracket #1 \rrbracket}
\newcommand\tgg[0]{\text{gg}}

\newcommand\urho[0]{{\underline{\rho}}}
\newcommand\uk[0]{{\underline{k}}}

\begin{document}

\title{\vspace{1.5cm}\bf
Nested ansatz method for Baker-Akhiezer functions
}

\author{
A. Mironov$^{b,c,d,}$\footnote{mironov@lpi.ru,mironov@itep.ru},
A. Morozov$^{a,c,d,}$\footnote{morozov@itep.ru},
A. Popolitov$^{a,c,d,}$\footnote{popolit@gmail.com}
}

\date{ }

\maketitle

\vspace{-6cm}

\begin{center}
  \hfill MIPT/TH-03/26\\
  \hfill FIAN/TD-03/26\\
  \hfill ITEP/TH-03/26\\
  \hfill IITP/TH-03/26
\end{center}

\vspace{4.5cm}

\begin{center}
$^a$ {\small {\it MIPT, Dolgoprudny, 141701, Russia}}\\
$^b$ {\small {\it Lebedev Physics Institute, Moscow 119991, Russia}}\\
$^c$ {\small {\it NRC ``Kurchatov Institute", 123182, Moscow, Russia}}\\
$^d$ {\small {\it Institute for Information Transmission Problems, Moscow 127994, Russia}}
\end{center}

\vspace{.1cm}

\begin{abstract}
  We explain that the logic behind the derivation of the Noumi-Shiraishi function can be
  applied directly to the Baker-Akhiezer function (BAF). This amounts to changing
  an ansatz for BAF to a nested one, where the BAF of $N+1$ variables is recursively expressed
  as a sum over BAFs of $N$ variables.
  This may be seen as a generalization of symmetrization trick from
  \cite{paper:MMP-twisted-cher}, but for the generally non-symmetric BAF.
  We demonstrate that, for usual non-twisted
  ($a=1$) BAFs, this method correctly reproduces the Noumi-Shiraishi formula
  directly from linear equations, resolving the ambiguity related to non-simple roots.
  For the first non-trivial twisted case ($N=3,\ a=2$) this method also fixes
  this ambiguity, moreover, answers for the first few layers of coefficients
  are in the form of \textit{direct quantization} of \cite{paper:MMP-twisted-cher}.
\end{abstract}

\bigskip

\newcommand\smallpar[1]{
  \noindent $\bullet$ \textbf{#1}
}

\section{Introduction}\label{sec:introduction}

Multivariable Baker-Akhiezer functions (BAFs), introduced by O. Chalykh
in \cite{paper:Ch-baf-original,paper:ES-alg-int}
are peculiar common polynomial eigenfunctions of the
Ruijsenaars-Schneider (RS) Hamiltonians possessing the bi-spectral symmetry.
These latter Hamiltonians were recently realized
\cite{paper:MMP-dim-comm-fams}
to be just one of infinitely many
integrable systems, each corresponding to a ray of operators in the Ding-Iohara-Miki
(DIM) algebra \cite{paper:DIM,paper:Miki}
with a certain slope $(a_1, a_2)$. The story actually has
simpler incarnations at the level of the $W_{1+\infty}$
\cite{paper:PRS, paper:FKRW, paper:AFMO, paper:KR-transformation-groups}
and affine Yangian algebras
\cite{paper:Ts-yangian, paper:P-yangian},
where it is most naturally related to superintegrable partition functions of
matrix models
\cite{paper:MOP-beta-models-directly,
  paper:MOP-beta-wlzz,
  paper:MMMP-comm-fams}. Still, for BAFs most natural seems to be the fully $(q,t)-$deformed level
of DIM algebra, where its connection to triad structure \cite{paper:MMP-basic-triad}
(interrelating BAFs \cite{paper:Ch-baf-original,ChE}, Macdonald polynomials \cite{Mac} and Noumi-Shiraishi (NS) functions \cite{paper:NS-direct-approach-to-bispectral})
 is most clearly seen.

At this level, Hamiltonian systems for other (distinct from RS) rays $(a_1,a_2)$ are solved
by so-called twisted BAFs \cite{ChE,paper:ChF-baf-twisted}
which are, in turn, expected to belong to their
respective $(a_1,a_2)$-triads, interrelating them with proper $(a_1,a_2)$ analogs of the
Macdonald polynomials and Noumi-Shiraishi functions.

However, manifest expressions for the $(a_1,a_2)$-triads are not known
(apart from the simplest $N=2$, $a_1=1$, $a_2 = a = 2$ case)
\cite{paper:MMP-chalykh-approach-to-dim-eigenfunctions},
the main
complication being the gauge freedom in the definition of (twisted) BAF which,
if not resolved correctly, completely obfuscates the triad structure.
In this letter, we propose a method to fix this gauge freedom in a particular way
that is consistent with the non-twisted ($a=1$) case, with help of the \textbf{nested ansatz}.

\bigskip

In the original formulation by O. Chalykh \cite{paper:Ch-baf-original},
the Baker-Akhiezer function is defined as a peculiar ansatz
\begin{align} \label{eq:baf-ansatz}
  \Psi\br{\uz,\ulambda}
  = q^{(\ulambda + m\urho, \uz)} \sum_{k_{ij}=0}^m q^{-\sum_{i<j}k_{ij}(z_i-z_j)}
  \psi_{\uk}\br{\ulambda}
\end{align}
satisfying a set of linear equations (periodicity conditions \cite{paper:MMPZ-ell-triad})
\begin{align} \label{eq:baf-periodicity-conditions}
  \Psi\br{z_k + j,\ulambda} = \Psi\br{z_l - j,\ulambda},\ j=1..m,
  \ \ k < l, \ \ z_k = z_l
\end{align}
He proves that such defined $\Psi\br{\uz,\ulambda}$ exists and is unique up to
normalization, as a function of $z$ and $\lambda$.

\bigskip

Inherent in this definition is ambiguity in choosing individual coefficients $\psi_{\uk}$.
This can be seen for the first time when $N=3$, $m=1$. In this case, there are two simple positive roots: $\alpha_{1,2}$
and $\alpha_{2,3}$, and one non-simple root $\alpha_{1,3} = \alpha_{1,2}+\alpha_{2,3}$,
therefore, the combination of coefficients $\tilde{\psi}_{1,1,0} = \tilde{\psi}_{1,1,0} + \tilde{\psi}_{0,0,1}$ enters all equations
as a monolithic sum. Hence, the function $\Psi$ is unambiguously
defined, while the splitting of $\widetilde{\psi}_{\uk}$ into individual $\psi_{\uk}$ is
essentially free.

In \cite{paper:MMP-chalykh-approach-to-dim}, we demonstrated that, in first simple cases,
it is possible to resolve this ambiguity in a smart way by \textit{additionally} requiring
certain pole structure of individual items $\psi_\uk$.
In \cite{paper:MMP-basic-triad}, this initial observation further led to discovery
of a remarkable \textit{triad structure} that interlinks BAFs, Noumi-Shiraishi functions
and Macdonald polynomials.

However, besides the triad relation to the NS function, which does have nicely factorized split
formula for its coefficients, there seemed to be no reason in general why BAF
groups of coefficients $\tilde{\psi}$ should split into nicely factorized
$q-$Pochhammer expressions. At the same time, case analysis we performed in
\cite{paper:MMP-twisted-cher} suggests that this \textit{is} the desired decomposition,
as it is the one consistent with the \textit{direct quantization} paradigm.

\bigskip

As we briefly explained in \cite{paper:MMP-baf-as-eigens}, and now rehearse
in Section~\ref{sec:ns-factorization}, the technical reason behind the factorization
of Macdonald (and NS) coefficients is that one is able to superimpose two independent
structures:
\begin{itemize}
\item one is the root lattice, consisting of $\alpha_{i,j} = e_i - e_j$ and, correspondingly,
  ratios $\br{\frac{x_i}{x_j}}$, $i < j$
\item another is nesting of permutation groups $S_N \supset S_{N-1} \supset ...$
  and, correspondingly, \textit{individual} $x_i, i=1..N$.
\end{itemize}

A natural question is, then, whether this superimposition of structures can be
performed directly on the level of Baker-Akhiezer function, i.e. on the level of the
corresponding ansatz and linear equation. This should be useful in situations where
triadic picture is not fully worked out yet: twisted BAFs being one main example,
and triads for other (than $A_{N-1}$) root systems being another.

\textbf{The claim} of this short letter is that this is, indeed, possible. We explain
how this works in case of usual BAF in Section~\ref{sec:nested-usual-baf} and
show that, indeed, the NS expansion is correctly reproduced this way.

In Section~\ref{sec:nested-twisted-baf}, we apply this reasoning to the first non-trivial twist
case, $a=2$. Already known (see \cite{paper:MMP-chalykh-approach-to-dim} eqn.(49))
is the $N=2$, $a=2$ twisted
BAF for an arbitrary $m$, and, in this Section, we use the nested ansatz method to get
manifest expressions
for the first few layers of coefficients in the $N=3$, $a=2$ case. Unfortunately, solving
even the nested linear system completely for high enough $m$
(which is needed to restore/guess formulas for generic $m$, see eqns.\eqref{eq:pol-01}--\eqref{eq:pol-81}) proves to be computationally challenging;
difficulty being all the more exacting since here
(in variance with \cite{paper:MMP-twisted-cher})
we are interested in the answers for arbitrary $\ulambda$, and not just at the
special point $\lambda = (0,m,\dots,(N-1)m)$.

Still, even the answers that \textit{are} available, demonstrate the tip of the emerging
structure:
\begin{itemize}
\item just as in the $N=2$, $a=2$ case, the answers are sums of fully factorized
  $q$-Pochhammer-like expressions
  (which are similar to the ones appearing in original Noumi-Shiraishi function);
\item dependence on $\Lambda_i = q^\lambda_i$ is fully encapsulated (hidden)
  inside peculiar homogeneous combinations $g_{i,j}(k)$ (see \eqref{eq:gg12-def});
\item apart from $g_{i,j}(k)$ factors, all other factors are symmetric $q$-numbers,
  where $m$ enters linearly (see, for instance, \eqref{eq:tpsi-1-1});
\item these two latter properties combined mean that the answers adhere to the general
  spirit of \textit{direct quantization}, advocated in \cite{paper:MMP-twisted-cher}.
  Moreover, answers analytically continue to arbitrary $t=q^{-m}$ straightforwardly.
\end{itemize}

These notable properties of the answers give some merit to the method they were obtained by:
to \textbf{the nested ansatz method}. Demonstrating usefulness and advantages of this method
is the main goal of this short letter.
We conclude in Section~\ref{sec:conclusion} with some plausible directions for future
applications.

\bigskip

\bigskip

\paragraph{Notation} We denote with an underline the sequences of variables,
for instance
\begin{align}
  \ux = x_1,\dots,x_N; \ \ \ \ulambda = \lambda_1, \dots, \lambda_N
\end{align}

$(x;q)_n$ is the $q$-Pochhammer symbol
\begin{align}
  (x;q)_n = \prod_{k=0}^{n-1} \br{1 - q^k x} = \frac{(x;q)_\infty}{(q^n x;q)_\infty}
\end{align}

$[n]$ and $[n]!$ (and, respectively, $\ssbr{n}$, $\ssbr{n}!$)
are the $q$-numbers and $q$-factorials in usual (and, respectively, symmetric)
conventions
\begin{align}
  [n] = \frac{q^n - 1}{q-1}\ \ \ [n]! = \prod_{k=1}^n [k];
  \ \ \ \   \ssbr{n} = \frac{q^{n/2} - q^{-n/2}}{q^{1/2}-q^{-1/2}}\ \ \
  \ssbr{n}! = \prod_{k=1}^n \ssbr{k}
\end{align}

\section{Nested ansatz for Macdonald polynomials and NS function}
\label{sec:ns-factorization}

Let us recall the reasoning behind original Macdonald's explicit formula
\cite{Mac}[eqns.(6.20),(6.24),(7.14’)].
and its Noumi-Shiraishi generalization \cite{paper:NS-direct-approach-to-bispectral}[eqn.(1.10,1.11)] to the case of non-integer partitions $\lambda$.

Key observation is that it is possible to expand Macdonald polynomial of $N$ variables
$x_1, \dots, x_N$ into Macdonald polynomials of $N-1$ variables $x_2,\dots,x_N$, with
$x_1$-dependent coefficients
\begin{align} \label{eq:mac-expansion}
  M_\lambda\br{x_1,\dots,x_N} = \sum_\mu c_{\lambda,\mu}(x_1) M_\mu\br{x_2,\dots,x_N}
\end{align}
where the expansion coefficients $c_{\lambda,\mu}(x_1)$ are nothing but
one-variable skew Macdonald polynomials
\begin{align} \label{eq:onevar-skew-macdonald}
  c_{\lambda,\mu}(x_1) = M_{\lambda/\mu}(x_1) =
  \prod_{1\leq i<j \leq n}
  \frac{\br{q^{\mu_j-\lambda_i+1} t^{j-i-1};q}_{\lambda_i-\mu_i}}
    {\br{q^{\mu_i - \lambda_j} t^{j-i};q}_{\lambda_i - \mu_i}}
  \prod_{1\leq i\leq j < n}
  \frac{\br{q^{\mu_i-\mu_j}t^{j-i+1};q}_{\lambda_i-\mu_i}}
    {\br{q^{\mu_i-\mu_j+1}t^{j-i};q}_{\lambda_i - \mu_i}}
  \cdot x_1^{|\lambda|-|\mu|}
\end{align}
Here, crucially, the formula is very explicit and manifest in terms of $\lambda$,$\mu$ and $x_1$.

The observation of Noumi-Shiraishi is then that the formula \eqref{eq:onevar-skew-macdonald}
makes sense not only for integral $\lambda \in \mathbb{Z}_{>0}$,
but for arbitrary complex $\lambda$. Correspondingly, the Noumi-Shiraishi function
(which now features infinite summation over the space of upper-triangular
$N$ by $N$  matrices $\mathcal{M}_N$, with entries $\theta_{i,j}$)
\begin{align} \label{eq:ns-def}
  \mathfrak{P}_{q,t}\br{\ux,\ulambda} = & \
  \prod_{i=1}^N\br{x_i^{\lambda_i - \beta \rho_i}}
  \sum_{\theta_{ij}=0}^\infty
  \Bigg{(}
  \prod_{n=2}^N \prod_{1\leq i < n} \prod_{s=0}^{\theta_{in}-1}
  \frac{\br{1 - q^{s+1-\theta_{in}} t^{-1}}
    \br{1 - t q^{s+\lambda_n-\lambda_i + \sum_{b>n}\br{\theta_{ib}-\theta_{nb}}}}
  }{\br{1 - q^{s-\theta_{in}}} \br{1 - q^{s+1+\lambda_n-\lambda_i
        + \sum_{b > n}\br{\theta_{ib}-\theta_{nb}}}}}
  \\ \notag
  & \quad\quad\quad\quad\quad\quad\quad\quad\quad
  \times
  \prod_{n=2}^N \prod_{1\leq i<j<n} \prod_{s=0}^{\theta_{in}-1}
  \frac{\br{1 - t q^{s+\lambda_j-\lambda_i + \sum_{b>n}\br{\theta_{ib}-\theta_{jb}}}}
    \br{1 - t^{-1} q^{s + 1 + \lambda_j - \lambda_i - \theta_{jn}
        +\sum_{b>n}\br{\theta_{ib}-\theta_{jb}}}}}
  {\br{1 - q^{s+1+\lambda_j-\lambda_i+\sum_{b>n}\br{\theta_{ib}-\theta_{jb}}}}
    \br{1-q^{s+\lambda_j-\lambda_i-k_{jn} + \sum_{b>n}\br{\theta_{ib}-\theta_{jb}}}}}
  \Bigg{)}
  \\ \notag
  & \quad\quad\quad\quad\quad\quad\quad\quad
  \prod_{1\leq i<j\leq N} \br{\frac{x_j}{x_i}}^{\theta_{ij}},
\end{align}
$\beta = \log_q(t)$,
is shown in \cite{paper:NS-direct-approach-to-bispectral} to be the simultaneous
eigenfunction of all Ruijsenaars-Schneider operators.
In terms of generating function for the RS operators this can be stated as follows
\begin{align}
  D^x(u) \mathfrak{P}_{q,t}\br{\ux,\ulambda} =
  \mathfrak{P}_{q,t}\br{\ux,\ulambda} \cdot \prod_{i=1}^N \br{1-u s_i}
\end{align}
Crucially, the whole line-of-thought builds upon on the possibility of expansion
\eqref{eq:mac-expansion} -- which is possible due to Macdonald polynomials being the basis
in the space of symmetric polynomials in \textit{any} number of variables $x$.

\section{Nested ansatz for non-twisted BAF} \label{sec:nested-usual-baf}

In \cite{paper:MMP-basic-triad} we discovered that O.Chalykh's Baker-Akhiezer function
\eqref{eq:baf-ansatz},\eqref{eq:baf-periodicity-conditions}
while defined from completely different considerations,
has striking relation to the Noumi-Shiraishi function
\eqref{eq:ns-def}
\begin{align}
  \Psi\br{\ux,\ulambda} = & \ \mathcal{N}_\ulambda \mathfrak{P}_{q,q^{-m}}\br{\ux,\ulambda}
  \\ \notag
  \mathcal{N}_\ulambda = & \ \prod_{k>l}\prod_{j=1}^m
  \br{q^{\frac{\lambda_k-\lambda_l}{2} - j} - q^{-\frac{\lambda_k-\lambda_l}{2}}}
\end{align}
which, furthermore, is a part of more complex triad structure, interrelating
BAFs, Macdonald polynomials and NS functions \cite[fig.1]{paper:MMP-basic-triad}.
\footnote{Note that the coefficient $\mathcal{N}_\ulambda$ occurs when
one is interested in bi-spectral dual $\Psi$:
$\Psi\br{\ux,\ulambda} = \Psi\br{\ulambda,\ux}$. In practice one often works
in non-symmetric normalization where the first coefficient of $\Psi$ is unity:
$\psi_{\underline{0}}(\ulambda) = 1$. In such normalization on simply has
$\Psi\br{\ux,\ulambda} = \mathfrak{P}_{q,q^{-m}}\br{\ux, \ulambda}$.}

In this triadic logic, a natural question is then, whether the reasoning
from the previous section, that relies on the decomposition into basis in
$x_2,\dots,x_N$, which is present for Macdonald and NS ``vertices'' of the triad,
may be somehow adapted and applied directly to the BAF ``vertex''.

Recall that (non-twisted) Baker-Akhiezer function is defined
(following \cite{paper:Ch-baf-original}) to be a function of variables
$x_1,\dots,x_N$ and $\lambda_1,\dots,\lambda_N$ of the form
\begin{align}\label{eq:baf-ansatz1}
  \Psi_m^{[N]}\br{\uz,\ulambda}
  = q^{(\ulambda + m\urho, \uz)} \sum_{k_{ij}=0}^m q^{-\sum_{i<j}k_{ij}(z_i-z_j)}
  \psi_{\uk}\br{\ulambda}, \ \ \ q^{z_i} = x_i
\end{align}
where we have explicitly specified $m$ and $N$ dependence of $\Psi$,
that satisfies a set of linear difference equations
\begin{align}\label{eq:baf-linear-equations1}
  \Psi_m^{[N]}\br{z_k + j,\ulambda} = \Psi_m^{[N]}\br{z_l - j,\ulambda},\ j=1..m,
  \ \ k < l, \ \ z_k = z_l
\end{align}
\noindent Let us try to apply the previous section logic to these circumstances.

\bigskip

Observe that when going
from $N-1$ variables to $N$ variables large portion of equations
\eqref{eq:baf-linear-equations1}
stays the same (with suitable change $x_i \rightarrow x_{i+1}$
, $\lambda_i\rightarrow\lambda_{i+1}$) and added are only the equations for roots
$\alpha_{1,k}, k=2..N$.

It makes sense then to decompose $\Psi_m^{[N]}(\ux,\ulambda)$ as a sum over
$\Psi_m^{[N-1]}(\ux,\underline{\mu})$ with suitable $\mu$: then
the ``old'' equations for roots $\alpha_{i,j},\ \ i < j=2..N$ would automatically hold.

Accurate degree counting of $x_i$ and $x_i^{\lambda_i}$ in the original ansatz
\eqref{eq:baf-ansatz} shows that the nested ansatz for BAF should be
\begin{align}\label{eq:baf-refined-ansatz}
  \Psi_m^{[N]}\br{x_1,\dots,x_N|\lambda_1,\dots,\lambda_N} = & \
  \\ \notag
  x_1^{\lambda_1 + m (N-1)/2}
  & \
  \sum_{k_{12},\dots,k_{1N}=0}^{m}
  \frac{\hpsi_{m,k_{12},\dots,k_{1N}}}
       {x_1^{\sum_{i=2}^N k_{1i}}}
       \Psi_m^{[N-1]}\br{x_2,\dots,x_N|\lambda_2 + k_{12} - \frac{m}{2},\dots,
         \lambda_N + k_{1N} - \frac{m}{2}}
\end{align}
that is, we sum over $\umu = \ulambda + \uDelta$ where the integral shift $\uDelta$
compensates for the power of $x_1$ in each item.

\bigskip

Starting now with $N=2$ BAF (for which directly solving linear equations
\eqref{eq:baf-linear-equations1} with original ansatz \eqref{eq:baf-ansatz1} has no ambiguity
-- but it similarly can be expressed as a nested ansatz over $\Psi^{[1]}(x,\lambda)$)
\begin{align}
  \Psi_m^{[2]}\br{x_1,x_2|\lambda_1,\lambda_2} =
  & \
  x_1^{\lambda_1 + m/2}
  \sum_{k_{12}=0}^{m}
  \frac{\hpsi_{m,k_{12}}}
       {x_1^{k_{12}}}
       \cdot
    \underbrace{x_2^{\lambda_2 + k_{12}-m/2}}_{
      \Psi_m^{[1]}\br{x_2|\lambda_2 + k_{12} - \frac{m}{2}}}
    \\ \notag
    \hpsi_{m,k_{12}} = & \
    (-1)^{k_{12}} q^{-m k_{12} + k_{12}(k_{12}-1)/2} \frac{[m]!}{[m-k_{12}]![k_{12}]!}
    \prod_{i=1}^{k_{12}} \frac{\br{\frac{\Lambda_1}{\Lambda_2} q^{m-i+1} - 1}}
         {\br{\frac{\Lambda_1}{\Lambda_2} q^{-i} - 1}};
         \ \ \Lambda_i = q^{\lambda_i}
\end{align}

In symmetric $q$-numbers this can be put into even nicer form
(which would continue to persist in the $a=2$ case, see \eqref{eq:psi-n1-0-homog-quant});
it is straightforward to see that this expression is equivalent to the NS formula
coefficients \eqref{eq:ns-def}
\begin{align}
  \hpsi_{m,k_{12}} = & \
  (-1)^{k_{12}} \frac{\ssbr{m}!}{\ssbr{m-k_{12}}!\ssbr{k_{12}}!}
  \prod_{i=0}^{k_{12}-1} \frac{\ssbr{\lambda_2-\lambda_1-m + i}}
  {\ssbr{\lambda_2-\lambda_1 + 1 + i}}
\end{align}

\noindent One can now solve the ``new'' equations for $\alpha_{1,k}, k=2..N$ for $N=3$
to manifestly and \textit{unambiguously} obtain
\begin{align} \label{eq:nested-ns-n=3}
  \hpsi_{m,k_{12},k_{13}} = & \
  (-1)^{k_{12}+k_{13}}
  \frac{\ssbr{m}!}{\ssbr{m-k_{12}}!\ssbr{k_{12}}!}
  \frac{\ssbr{m}!}{\ssbr{m-k_{13}}!\ssbr{k_{13}}!}
  \prod_{i=0}^{k_{12}-1} \frac{\ssbr{\lambda_2-\lambda_1 + i - m}}
       {\ssbr{\lambda_2-\lambda_1 + i+1}}
       \prod_{i=0}^{k_{13}-1} \frac{\ssbr{\lambda_3-\lambda_1 + i - m}}
       {\ssbr{\lambda_3-\lambda_1 + i+1}}
       \\ \notag
       & \quad\quad\quad\quad\quad\quad\quad
       \times \prod_{i=0}^{k_{13}-1}
       \frac{\ssbr{\lambda_3-\lambda_2 + i - m} \ssbr{\lambda_3-\lambda_2 + i + m + 1 -k_{12}}}
            {\ssbr{\lambda_3-\lambda_2 + i+1} \ssbr{\lambda_3-\lambda_2 + i-k_{12}}}
\end{align}
\noindent Similarly explicit and fully factorized formula takes place for $N=4$,
that is for $\hpsi_{m,k_{12},k_{13},k_{14}}$. Observe how the first line
in \eqref{eq:nested-ns-n=3} directly corresponds to the first line
of \eqref{eq:ns-def} and the second one, respectively, to the second.
One just has to perform the trivial change of $\lambda$-variables
$\lambda_i \leftrightarrow -\lambda_{N-i+1}$,
which, in other words, corresponds to $x_N$, and not $x_1$, being the distinguished
variable in \eqref{eq:baf-refined-ansatz}.

To summarize, \textbf{the Noumi-Shiraishi formula is correctly reproduced}
with help of nested ansatz.

\section{Nested ansatz for twisted BAF} \label{sec:nested-twisted-baf}

Having checked the consistency of our method in the non-twisted $a=1$ case, we now turn
our attention to the more interesting case of twisted BAFs (tBAFs)

As a base for our method we have the previously obtained $N=2$ $a=2$ tBAF
\begin{align} \label{eq:baf-twisted-base}
  \Psi_m^{(2)[2]}\br{x_1,x_2|\lambda_1,\lambda_2} = & \ \
  x_1^{\lambda_1} x_2^{\lambda_2} \cdot x_1^m x_2^{-m}
  \cdot \sum_{k=0}^{2 m}
  \psi^{(2)[2]}_m(\lambda_1,\lambda_2) \cdot   \br{\frac{x_2}{x_1}}^k
  \\ \notag
  \psi^{(2)[2]}_m(\lambda_1,\lambda_2) = & \
  \Bigg{(}
    \sum_{r=\text{max}(0,k-m)}^{\lfloor \frac{k}{2} \rfloor}
    (-1)^{r}
    q^{
      (k - 2 r)(-m + \frac{1}{2} k)
      + \frac{1}{2} r (r + 1)
    }
    (1-q)^{k - 2 r}
    \frac{\sbr{m+k-2 r}!}{\sbr{m-k+r}!\sbr{k-2 r}! \sbr{r}!}
    \\ \notag
    & \quad \quad \quad \quad
    \br{\Lambda_1 \Lambda_2}^{k - 2 r}
    \frac{\prod_{i=1}^r \text{ff}_{1,2} (i - 1 - m) \prod_{i=0}^{r-1} \text{ff}_{1,2}(k-i)}{\prod_{i=1}^k \text{ff}_{1,2}(i)}
  \Bigg{)}
\end{align},
where $x_i = q^{z_i/a}$ and $\Lambda_i = q^{\lambda_i}$ -- as compared to the original
Chalykh's definition. The peculiar factor is
\begin{align}
  \tff_{i,j}(k) = \br{\Lambda_i^a - q^k \Lambda_j^a}
\end{align}
\noindent The analog of nested ansatz \eqref{eq:baf-refined-ansatz} reads

\begin{empheq}[box=\fbox]{align} \label{eq:baf-twisted-ansatz}
  \Psi_m^{(a)[N]}\br{x_1,\dots,x_N|\lambda_1,\dots,\lambda_N} = & \
  x_1^{\lambda_1 + a m (N-1)/2}
  \sum_{k_{1,2},\dots,k_{1,N}=0}^{m a}
  \frac{\tilde{\psi}^{(a)}_{m,k_{1,2},\dots,k_{1,N}}}
       {x_1^{\sum_{i=2}^N k_{1,i}}}
       \\ \notag
       & \ \ \ \Psi_m^{(a)[N-1]}\br{x_2,\dots,x_N|\lambda_2 + k_{1,2} - \frac{m a}{2},\dots,\lambda_N + k_{1,N} - \frac{m a}{2}}
\end{empheq}

\noindent and the linear equations are (only the ones new to $N$ variables)
\begin{empheq}[box=\fbox]{align}\label{eq:baf-twisted-lineqs}
  \Psi^{(a)[N]}_m \br{x_i \epsilon q^{-\frac{j}{2 a}},
    \dots,
    x_i q^{\frac{j}{2 a}}, \dots}
  = \epsilon^j \Psi^{(a)[N]}_m \br{x_i \epsilon q^{\frac{j}{2 a}},
    \dots,
    x_i q^{-\frac{j}{2 a}}, \dots}
\end{empheq}

\subsection{Peculiarities of the $N=3$ $a=2$ case}
The simplest non-trivial case, where nested ansatz is applicable,
is the $N=3$, $a=2$ case. Here we discuss the structure of answers that already
starts to be visible in this case\footnote{At present, even in the $N=3,\ a=2$
we are unable to obtain all answers for \textit{generic} $m$. Still, the answers we can
obtain have peculiar structure, which we want to describe.
To obtain general $m$ answers and to move to $N>3$ and/or $a>2$ cases arguably requires
some crossing of ideas from the present paper and \cite{paper:MMP-twisted-cher}
-- which should be possible since answers seem to adhere to direct quantization prescription.
}.

The $N=3$ unknown coefficients have two indices: corresponding to
multiplicities of roots $\alpha_{1,2}$ and $\alpha_{1,3}$.

As in \eqref{eq:baf-twisted-base} some coefficients are factorized, and some are not,
and we describe them now in the order of increasing complexity.

The first non-factorizable coefficient is $\tpsi_{m,2,0}$. For subsequent $m$ it is equal to
\begin{align}
  \tpsi_{1,2,0} = & \ \frac{[m]}{\tff_{1,2}(1) \tff_{1,2}(2)}
  \cdot (-q) \cdot \tff_{1,2}(-1) \tff_{1,2}(2)
  \\ \notag
  \tpsi_{2,2,0} = & \ \frac{[m]}{\tff_{1,2}(1) \tff_{1,2}(2)}
  \cdot
  \br{(-q) \cdot \tff_{1,2}(-2) \tff_{1,2}(2) + \frac{[4][3] (q-1)^2}{[2]q^2}
    \br{\Lambda_1\Lambda_2}^2
  }
  \\ \notag
  \tpsi_{3,2,0} = & \ \frac{[m]}{\tff_{1,2}(1) \tff_{1,2}(2)}
  \cdot
  \br{(-q) \cdot \tff_{1,2}(-3) \tff_{1,2}(2) + \frac{[4][5] (q-1)^2}{q^4}
    \br{\Lambda_1\Lambda_2}^2
  }
  \\ \notag
  \tpsi_{4,2,0} = & \ \frac{[m]}{\tff_{1,2}(1) \tff_{1,2}(2)}
  \cdot
  \br{(-q) \cdot \tff_{1,2}(-4) \tff_{1,2}(2) + \frac{[6][5][3] (q-1)^2}{[2]q^6}
    \br{\Lambda_1\Lambda_2}^2
  }
  \\ \notag
  \tpsi_{5,2,0} = & \ \frac{[m]}{\tff_{1,2}(1) \tff_{1,2}(2)}
  \cdot
  \br{(-q) \cdot \tff_{1,2}(-5) \tff_{1,2}(2) + \frac{[6][7][4] (q-1)^2}{[2] q^8}
    \br{\Lambda_1\Lambda_2}^2
  },
\end{align}
where from, in general, we conclude
\begin{align}
  \tpsi_{m,2,0} = & \ \frac{[m]}{\tff_{1,2}(1) \tff_{1,2}(2)}
  \cdot
  \br{(-q) \cdot \tff_{1,2}(-m) \tff_{1,2}(2) + \frac{[m-1][m+1][m+2] (q-1)^2}{[2]q^{2(m-1)}}
    \br{\Lambda_1\Lambda_2}^2
  }
\end{align}
which is \textit{in accordance} with the structure, of \eqref{eq:baf-twisted-base}.

\bigskip

Going through similar motions, one obtains for $\tpsi_{m,n_1,0}$
\begin{align} \label{eq:psi-n1-0}
  \tpsi_{m,n_1,0} = & \
    \Bigg{(}
    \sum_{r=\text{max}(0,n_1-m)}^{\lfloor \frac{n_1}{2} \rfloor}
    (-1)^{r}
    q^{\frac{n_1^2}{2} + \frac{r^2}{2} - n_1 r +\frac{r}{2}
    }
    q^{-(n_1-2 r) m}
    (1-q)^{n_1 - 2 r}
    \frac{\sbr{m+n_1-2 r}!}{\sbr{m-n_1+r}!\sbr{n_1-2 r}! \sbr{r}!}
    \\ \notag
    & \quad \quad \quad \quad
    \br{\Lambda_1 \Lambda_2}^{n_1 - 2 r}
    \frac{\prod_{i=1}^r \text{ff}_{1,2} (i - 1 - m) \prod_{i=0}^{r-1} \text{ff}_{1,2}(n_1-i)}{\prod_{i=1}^{n_1} \text{ff}_{1,2}(i)}
  \Bigg{)}
\end{align}
that is, at the end of the day the formula is \textit{literally} the same as
\eqref{eq:baf-twisted-base}, with trivial change $\Lambda_{2,3} \rightarrow \Lambda_{1,2}$.

\paragraph{Dequantization} This formula \eqref{eq:psi-n1-0} can be, in fact,
made much simpler by, firstly, using \textit{symmetric} $q$-numbers
\begin{align} \label{eq:qnumsym-def}
  \llbracket n \rrbracket := \frac{q^{n/2} - q^{-n/2}}{q^{1/2} - q^{-1/2}}
\end{align}
and, secondly, considering \textit{symmetric} root-related quantities instead of
not quite symmetric $\tff_{1,2}(k)$
\begin{align} \label{eq:gg12-def}
  \tgg_{1,2}(k) := \frac{q^{-k/2} \tff_{1,2}(k)}{(q^{-1/2} - q^{1/2}) \Lambda_1\Lambda_2}
  = \frac{\br{q^{-k/2}\frac{\Lambda_1}{\Lambda_2} - q^{k/2} \frac{\Lambda_2}{\Lambda_1}}}
  {\br{q^{-1/2} - q^{1/2}}} \mathop{=}_{\Lambda_l = q^{\lambda_l/2}}
  \llbracket \lambda_2 - \lambda_1 + k \rrbracket
  =_{q \rightarrow 1}
  \br{\lambda_2 - \lambda_1 + k}
\end{align}

With help of these two devices the formula becomes (note that explicit $\Lambda_l$ entirely disappears from the formula -- only the ``root-related'' quantities $\tgg_{i,j}(k)$ enter)
\begin{align} \label{eq:psi-n1-0-homog-quant}
  \eqref{eq:psi-n1-0} = & \
  \Bigg{(}
    \sum_{r=\text{max}(0,n_1-m)}^{\lfloor \frac{n_1}{2} \rfloor}
    (-1)^{r}
    \frac{\ssbr{m+n_1-2 r}!}{\ssbr{m-n_1+r}!\ssbr{n_1-2 r}! \ssbr{r}!}
    \frac{\prod_{i=1}^r \tgg_{1,2} (i - 1 - m) \prod_{i=0}^{r-1} \tgg_{1,2}(n_1-i)}{\prod_{i=1}^{n_1} \tgg_{1,2}(i)}
  \Bigg{)}
\end{align}
moreover there are no \textit{ad hoc} $q$-factors;
The \textit{classical limit} $q \rightarrow 1$ is, accordingly, very simple
\begin{align} \label{eq:psi-n1-0-class}
  \eqref{eq:psi-n1-0} \mathop{=}_{\substack{\Lambda_l = q^{\lambda_l/2} \\ q\rightarrow 1}}
  & \
  \Bigg{(}
  \sum_{r=\text{max}(0,n_1-m)}^{\lfloor \frac{n_1}{2} \rfloor}
  (-1)^{r}
  \frac{\br{m+n_1-2 r}!}{\br{m-n_1+r}!\br{n_1-2 r}! \br{r}!}
  \frac{\prod_{i=1}^r (\lambda_2 - \lambda_1 + i - 1 - m)
    \prod_{i=0}^{r-1} (\lambda_2 - \lambda_1 + n_1-i)}
       {\prod_{i=1}^{n_1} \br{\lambda_2 - \lambda_1 + i}}
       \Bigg{)}
\end{align}

The relation between \eqref{eq:psi-n1-0-class} and \eqref{eq:psi-n1-0-homog-quant}
is then, naturally a \textit{dequantization prescription} -- to be applied in more
complicated cases. Or, put another way, should one somehow obtain classical answer first,
representing it in the form \eqref{eq:psi-n1-0-class} allows for its straightforward
promotion to the full quantum answer \eqref{eq:psi-n1-0-homog-quant}. This is exactly
in spirit of \textit{direct quantization} paradigm, first signs of which are observed
in \cite{paper:MMP-twisted-cher}; now it receives further evidence.

\bigskip Slightly more complicated are the coefficients $\tpsi_{m,0,n_2}$, where
good basis functions start to get extra factors,
typical for Noumi-Shiraishi answers
(the ``core'' of the formula is still about the root $e_{1,3}$, but extra factors
are related to the ``old'' $N=2$ root $e_{2,3}$)
\begin{align} \label{eq:psi-0-n2}
  \tpsi_{m,0,n_2} = & \
    \Bigg{(}
    \sum_{r=\text{max}(0,n_2-m)}^{\lfloor \frac{n_2}{2} \rfloor}
    (-1)^{r}
    q^{\frac{n_2^2}{2} + \frac{r^2}{2} - n_2 r +\frac{r}{2}
    }
    q^{-(n_2-2 r) m}
    (1-q)^{n_2 - 2 r}
    \frac{\sbr{m+n_2-2 r}!}{\sbr{m-n_2+r}!\sbr{n_2-2 r}! \sbr{r}!}
    \\ \notag
    & \quad \quad \quad \quad
    \br{\Lambda_1 \Lambda_3}^{n_2 - 2 r}
    \frac{\prod_{i=1}^r \text{ff}_{1,3} (i - 1 - m) \prod_{i=0}^{r-1} \text{ff}_{1,3}(n_1-i)}{\prod_{i=1}^{n_2} \text{ff}_{1,3}(i)}
    \Bigg{)}
    \\ \notag
    & \ \quad\quad \times
    \br{\prod_{i=0}^{n_2 - 1} \frac{\tff_{2,3}\br{-m+i}}{\tff_{2,3}\br{i}}}
    \br{\prod_{i=1}^{n_2} \frac{\tff_{2,3}\br{m+i}}{\tff_{2,3}\br{i}}}
\end{align}

which, with help of similarly introduced $\tgg_{1,3}(k)$ and $\tgg_{2,3}(k)$ becomes
\begin{align} \label{eq:tpsi-0-n2}
  \tpsi_{m,0,n_2} = & \
    \Bigg{(}
    \sum_{r=\text{max}(0,n_2-m)}^{\lfloor \frac{n_2}{2} \rfloor}
    (-1)^{r}
    \frac{\ssbr{m+n_2-2 r}!}{\ssbr{m-n_2+r}!\ssbr{n_2-2 r}! \ssbr{r}!}
    \\ \notag
    & \quad \quad \quad \quad
    \frac{\prod_{i=1}^r \tgg_{1,3} (i - 1 - m) \prod_{i=0}^{r-1} \tgg_{1,3}(n_2-i)}{\prod_{i=1}^{n_2} \tgg_{1,3}(i)}
    \Bigg{)}
    \\ \notag
    & \ \quad\quad \times
    \br{\prod_{i=0}^{n_2 - 1} \frac{\tgg_{2,3}\br{-m+i}}{\tgg_{2,3}\br{i}}}
    \br{\prod_{i=1}^{n_2} \frac{\tgg_{2,3}\br{m+i}}{\tgg_{2,3}\br{i}}}
\end{align}
\noindent \textbf{Note} how the structure of this formula parallels the general
structure of $a=1$ Noumi-Shiraishi formulas \eqref{eq:ns-def}, \eqref{eq:nested-ns-n=3}:
the $\alpha_{2,3}$-root has the associated factors in the numerator, involving $m$
and $-m$, while $\alpha_{1,3}$-root has one factor with $-m$, one with $+1$, and
the $\lambda$-independent $q$-multinomial factor. This structure seems, therefore,
to be essential to the whole Noumi-Shiraishi construction, as it persists
to the twisted case here, and in more complicated examples below.
In future, the reason behind this phenomenon needs to be understood.

The homogeneous symmetric quantities \eqref{eq:qnumsym-def} and
\eqref{eq:gg12-def} prove to be the right
terms indeed, as the more complicated answers are equally simple. For instance
$\tpsi_{m,1,1}$ equals
\begin{align} \label{eq:tpsi-1-1}
  \tpsi_{m,1,1}
   = & \
   \ssbr{m}^2\ssbr{m+1}
   \frac{1}{\tgg_{1,2}(1)}\cdot\frac{\tgg_{1,3}(-m)}{1}
   \cdot\frac{1}{\tgg_{2,3}(-1)\tgg_{2,3}(1)}
   \\ \notag
   + & \ \ \ \   \ssbr{m}^2\ssbr{m+1}
   \frac{\tgg_{1,2}(-m)}{1} \cdot\frac{1}{\tgg_{1,3}(1)}
   \cdot\frac{1}{\tgg_{2,3}(-1)\tgg_{2,3}(1)}
  \\ \notag
  &
  + \ \ \ \ \ssbr{m}^3\ssbr{m+1}
  \quad\quad\quad\quad\quad\quad \frac{1}{\tgg_{1,2}(1)}\cdot\frac{1}{\tgg_{1,3}(1)}
  \cdot\frac{\tgg_{2,3}(-m-1)\tgg_{2,3}(m+1)}{\tgg_{2,3}(-1)\tgg_{2,3}(1)}
\end{align}
where now the answer stops being monomial in $\tgg_{2,3}(\bullet)$.

\paragraph{The R-operators} At this point it is convenient to introduce the following
``raising'' operators $R_{i,j}$, for the convenience of writing formulas like
\eqref{eq:tpsi-0-n2} and \eqref{eq:tpsi-1-1}. Namely, the operator $R_{1,2}$
(and, analogously, $R_{1,3}$) acts by successfully removing $\tgg_{1,2}(k)$
with the \textit{highest} $k$ from the denominator and putting
the corresponding $\tgg_{1,2}(-m + k_{max} - k)$ to the numerator.
For instance
\begin{align}
  \frac{1}{\tgg_{1,2}(1)\tgg_{1,2}(2)\tgg_{1,2}(3)}
  \mathop\rightarrow_{R_{1,2}} \frac{\tgg_{1,2}(-m)}{\tgg_{1,2}(1)\tgg_{1,2}(2)}
  \mathop\rightarrow_{R_{1,2}} \frac{\tgg_{1,2}(-m)\tgg_{1,2}(-m+1)}{\tgg_{1,2}(1)}
  \mathop\rightarrow_{R_{1,2}} \frac{\tgg_{1,2}(-m)\tgg_{1,2}(-m+1)\tgg_{1,2}(-m+2)}{1}
\end{align}

The operator $R_{2,3}$ is slightly more tricky.
It depends on additional shift $\delta$ and, ``determining'' the $\tgg_{2,3}$
with the lowest (negative) and highest (positive) arguments in the denominator,
multiplies the corresponding monomial by
\begin{align}
  \tgg_{2,3}(-m - k_{min} + \delta)\tgg_{2,3}(m + k_{max} - \delta)
\end{align}

\bigskip

With help of this notation $\tpsi_{m,0,n2}$ becomes simply
\begin{align}
  \tpsi_{m,0,n_2} = & \
  \br{\prod_{i=0}^{n_2 - 1} \frac{\tgg_{2,3}\br{-m+i}}{\tgg_{2,3}\br{i}}}
  \br{\prod_{i=1}^{n_2} \frac{\tgg_{2,3}\br{m+i}}{\tgg_{2,3}\br{i}}}
  \\ \notag
  & \quad\quad\quad\quad
  \Bigg{(}
  \sum_{r=\text{max}(0,n_2-m)}^{\lfloor \frac{n_2}{2} \rfloor}
  (-1)^{r}
  \frac{\ssbr{m+n_2-2 r}!}{\ssbr{m-n_2+r}!\ssbr{n_2-2 r}! \ssbr{r}!}
  R_{1,3}^r
  \Bigg{)} \frac{1}{\prod_{i=1}^{n_2} \tgg_{1,3}(i)}
\end{align}
while the more complicated $\tpsi_{m,1,1}$ is written as
\begin{align} \label{eq:tpsi-1-1-r}
  \tpsi_{m,1,1}
  = & \
  \Bigg{(}
  \ssbr{m}^2\ssbr{m+1} R_{1,2}
  + \ssbr{m}^2\ssbr{m+1} R_{1,3} + \ssbr{m}^3\ssbr{m+1} R_{2,3}(0)
  \Bigg{)}
  \frac{1}{\tgg_{1,2}(1)\tgg_{1,3}(1)\tgg_{2,3}(-1)\tgg_{2,3}(1)}
\end{align}
Note how direct quantization/dequantization continues to hold.

\paragraph{The general structure of $\tpsi_{m,n_1,n_2}$}

In general, looking at many specific examples, one can conclude that
$\tpsi_{m,n_1,n_2}$ has the following structure
\begin{align}
  \tpsi_{m,n_1,n_2} = \text{Pol}_{m,n_1,n_2}\br{R_{1,2},R_{1,3},R_{2,3}(\delta)}
  \frac{1}{\text{Denom}(m,n_1,n_2)}
\end{align}
where $\text{Pol}_{m,n_1,n_2} \br{R_{1,2},R_{1,3},R_{2,3}}$ is a certain
polynomial in symbols $R_{1,2}$, $R_{1,3}$, $R_{2,3}$ with $m$-dependent
coefficients (which are assembled from symmetric $q$-numbers).

$\text{Denom}(m,n_1,n_2)$ is simple and manifest expression
\begin{align}
  \text{Denom}(m,n_1,n_2) = \prod_{i=1}^{n_1} \tgg_{1,2}(i)
  \prod_{j=1}^{n_2} \tgg_{1,3}(j)
  \prod_{k=0}^{n_2-1} \br{\tgg_{2,3}(-n_1+k)\tgg_{2,3}(n_2-k)}
\end{align}

\paragraph{Answers for level $n_2=1$: $\tpsi_{m,n_1,1}$}

Clearly, the $\text{Denom}$ function in this case equals
\begin{align}
  \text{Denom}(m,n_1,1) = \br{\prod_{i=1}^{n_1} \tgg_{1,2}(i)}
  \cdot
  \tgg_{1,3}(1)
  \cdot
  \br{\tgg_{2,3}(-n_1)\tgg_{2,3}(1)}
\end{align}

And the corresponding operator polynomials $\text{Pol}_{m,n_1,1}$
turn out to be equal, in first few cases, to
\begin{align} \label{eq:pol-01}
  \text{Pol}_{m,0,1}
  = & \ssbr{m} \ssbr{m+1} R_{2,3}(0)
\end{align}
\begin{align}
  \text{Pol}_{m,1,1}
  = & \ssbr{m}^2 \ssbr{m+1} R_{1,2} + \ssbr{m}^2 \ssbr{m+1} R_{1,3}
  + \ssbr{m}^3 \ssbr{m+1} R_{2,3}(0)
\end{align}
\begin{align}
  \text{Pol}_{m,2,1}
  = &
  (-1) \ssbr{m}^2\ssbr{m+1} R_{1,2} R_{2,3}(1)
  \\ \notag
  + & \ \ssbr{m-1}\ssbr{m}^3\ssbr{m+1} R_{1,2}
  + \ssbr{m-1}\ssbr{m}^2\ssbr{m+1}^2 R_{1,3}
  + \frac{\ssbr{m-1}\ssbr{m}^3\ssbr{m+1}^2}{\ssbr{2}} R_{2,3}(0)
\end{align}
\begin{align}
  \text{Pol}_{m,3,1}
  = & (-1) \ssbr{m-1}\ssbr{m}^2\ssbr{m+1} R_{1,2}^2
  + (-1) \ssbr{m-1}\ssbr{m}^2\ssbr{m+1} R_{1,2} R_{1,3}
  + (-1) \ssbr{m-1}\ssbr{m}^3\ssbr{m+1} R_{1,2} R_{2,3}(1)
  \\ \notag
  + & \frac{\ssbr{m-2}\ssbr{m-1}\ssbr{m}^3\ssbr{m+1}^2}{\ssbr{2}} R_{1,2}
  + \frac{\ssbr{m-2}\ssbr{m-1}\ssbr{m}^2\ssbr{m+1}^2\ssbr{m+2}}{\ssbr{2}} R_{1,3}
  \\ \notag
  + & \frac{\ssbr{m-2}\ssbr{m-1}\ssbr{m}^3\ssbr{m+1}^2\ssbr{m+2}}{\ssbr{2}\ssbr{3}} R_{2,3}(0)
\end{align}
\begin{align}
  \text{Pol}_{m,4,1}
  = & \frac{\ssbr{m-1}\ssbr{m}^2\ssbr{m+1}}{\ssbr{2}} R_{1,2}^2 R_{2,3}(2)
  \\ \notag
  + &
  (-1) \ssbr{m-2}\ssbr{m-1}\ssbr{m}^3\ssbr{m+1} R_{1,2}^2
  + (-1) \ssbr{m-2}\ssbr{m-1}\ssbr{m}^2\ssbr{m+1}^2 R_{1,2} R_{1,3}
  \\ \notag
  + & \ (-1) \frac{\ssbr{m-2}\ssbr{m-1}\ssbr{m}^3\ssbr{m+1}^2}{\ssbr{2}} R_{1,2}R_{2,3}(1)
  \\ \notag
  + & \frac{\ssbr{m-3}\ssbr{m-2}\ssbr{m-1}\ssbr{m}^3\ssbr{m+1}^2\ssbr{m+2}}
  {\ssbr{2}\ssbr{3}} R_{1,2}
  + \frac{\ssbr{m-3}\ssbr{m-2}\ssbr{m-1}\ssbr{m}^2\ssbr{m+1}^2\ssbr{m+2}\ssbr{m+3}}
  {\ssbr{2}\ssbr{3}} R_{1,3}
  \\ \notag
  + & \ \frac{\ssbr{m-3}\ssbr{m-2}\ssbr{m-1}\ssbr{m}^3\ssbr{m+1}^2\ssbr{m+2}\ssbr{m+3}}
  {\ssbr{2}\ssbr{3}\ssbr{4}} R_{2,3}(0)
\end{align}

At this point we already see that the smaller $m$-dependent $q$-number
for $\tpsi_{m,n1,1}$ is $\ssbr{m - n1 + 1}$, therefore, for the following ones
we cannot really restore all the coefficients from our available data.

Some parts of the expression, however, \textit{can} be fixed. Namely,
\begin{align}
  \text{Pol}_{m,5,1}
  = & \frac{\ssbr{m-2}\ssbr{m-1}\ssbr{m}^2\ssbr{m+1}}{\ssbr{2}} R_{1,2}^3
  + \frac{\ssbr{m-2}\ssbr{m-1}\ssbr{m}^2\ssbr{m+1}}{\ssbr{2}} R_{1,2}^2 R_{1,3}
  + \frac{\ssbr{m-2}\ssbr{m-1}\ssbr{m}^3\ssbr{m+1}}{\ssbr{2}} R_{1,2}^2 R_{2,3}(2)
  \\ \notag
  + & \ (-1)\frac{\ssbr{m-3}\ssbr{m-2}\ssbr{m-1}\ssbr{m}^3\ssbr{m+1}^2}{\ssbr{2}} R_{1,2}^2
  + (-1)\frac{\ssbr{m-3}\ssbr{m-2}\ssbr{m-1}\ssbr{m}^2\ssbr{m+1}^2\ssbr{m+2}}{\ssbr{2}}
  R_{1,2} R_{1,3}
  \\ \notag
  + & \ (-1)\frac{\ssbr{m-3}\ssbr{m-2}\ssbr{m-1}\ssbr{m}^3\ssbr{m+1}^2\ssbr{m+2}}
  {\ssbr{2}\ssbr{3}}
  R_{1,2} R_{2,3}(1)
  \\ \notag
  + & \ssbr{m-4} \cdot \text{smth}_4
\end{align}
where we denote with $\text{smth}_4$ the piece we cannot restore/verify

Further,
\begin{align}
  \text{Pol}_{m,6,1}
  = & \ (-1)\frac{\ssbr{m-2}\ssbr{m-1}\ssbr{m}^2\ssbr{m+1}}{\ssbr{2}\ssbr{3}}
  R_{1,2}^3 R_{2,3}(3)
  \\ \notag
  + & \
  \frac{\ssbr{m-3}\ssbr{m-2}\ssbr{m-1} \ssbr{m}^3\ssbr{m+1}}{\ssbr{2}}
  R_{1,2}^3
  + \frac{\ssbr{m-3}\ssbr{m-2}\ssbr{m-1}\ssbr{m}^2\ssbr{m+1}^2}{\ssbr{2}}
  R_{1,2}^2 R_{1,3}
  \\ \notag
  + & \ \frac{\ssbr{m-3}\ssbr{m-2}\ssbr{m-1}\ssbr{m}^3\ssbr{m+1}^2}
  {\ssbr{2}^2}
  R_{1,2}^2 R_{2,3}(2)
  \\ \notag
  + & \ \ssbr{m-4} \cdot \text{smth}_4 + \ssbr{m-5} \ssbr{m-4} \cdot \text{smth}_5
\end{align}
\begin{align}
  \text{Pol}_{m,7,1}
  = &
  \ (-1)\frac{\ssbr{m-3}\ssbr{m-2}\ssbr{m-1}\ssbr{m}^2\ssbr{m+1}}{\ssbr{2}\ssbr{3}}
  R_{1,2}^4
  + (-1)\frac{\ssbr{m-3}\ssbr{m-2}\ssbr{m-1}\ssbr{m}^2\ssbr{m+1}}{\ssbr{2}\ssbr{3}}
  R_{1,2}^3 R_{1,3}
  \\ \notag
  + & \ (-1)\frac{\ssbr{m-3}\ssbr{m-2}\ssbr{m-1}\ssbr{m}^3\ssbr{m+1}}
  {\ssbr{2}\ssbr{3}}
  R_{1,2}^3 R_{2,3}(3)
  \\ \notag
  + & \ssbr{m-4} \cdot \text{smth}_4 + \ssbr{m-5} \ssbr{m-4} \cdot \text{smth}_5
  + \ssbr{m-6} \ssbr{m-5} \ssbr{m-4} \cdot \text{smth}_6
\end{align}
\begin{align} \label{eq:pol-81}
  \text{Pol}_{m,8,1}
  = & \frac{\ssbr{m-3}\ssbr{m-2}\ssbr{m-1}\ssbr{m}^4\ssbr{m+1}}
  {\ssbr{2}\ssbr{3}\ssbr{4}}
  R_{1,2}^4 R_{2,3}(4)
  \\ \notag
  + & \ssbr{m-4} \cdot \text{smth}_4 + \ssbr{m-5} \ssbr{m-4} \cdot \text{smth}_5
  + \ssbr{m-6} \ssbr{m-5} \ssbr{m-4} \cdot \text{smth}_6
  \\ \notag
  + & \ssbr{m-7} \ssbr{m-6} \ssbr{m-5} \ssbr{m-4} \cdot \text{smth}_7
\end{align}

\bigskip

All these answers continue to confirm the principle of direct quantization.
Moreover, observe, how generalization from integer $m$ to generic $t=q^{-m}$,
which is done separately for each $\ssbr{m+\bullet}$ and $\tgg_{i,j}(k)$
factor, is evident. At the moment we refrain from writing/conjecturing general formula
even for $\text{Pol}_{n_1, 1}$, and even more so for $\text{Pol}_{n_1, n_2}$
until we are able to gather more experimental data.
Still, even the presented partial answers illustrate the main point of the letter:
the nested ansatz method:
\begin{enumerate}
\item can be applied directly to (twisted) BAF;
\item leads to unambiguous answers (in the sense that splitting
  of $\tilde{psi}$ to $\hpsi$ is resolved);
\item which are manifestly a direct quantization, even in the stronger sense than
  that of \cite{paper:MMP-twisted-cher}, since no insertion of ad hoc
  $q^\bullet-$factors is needed: everything is accounted for by considering symmetric $q$-numbers.
\end{enumerate}

\section{Conclusion} \label{sec:conclusion}

In this letter, we presented the nested ansatz method for calculating Baker-Akhiezer
functions in the sense of O. Chalykh \cite{paper:Ch-baf-original} in various setups.
The method is particularly useful in situations when the triad picture is not fully established,
which we demonstrated by obtaining first few answers for coefficients of twisted BAF
in the first unknown non-trivial case of $N=3,\ a=2$.

Crucially, these answers continue to confirm the original educated guess
of \cite{paper:MMP-twisted-cher} that full quantum answers in the proper basis
are restored from their classical counterparts by simply changing numbers
to $q$-numbers, and Pochhammer symbols to $q$-Pochhammer symbols: the so-called
\textit{direct quantization}.

Generally speaking, the method is an improvement upon the initial ansatz
\eqref{eq:baf-ansatz}, and may be used with success in all situations where
analogs of BAF appear. Immediate and most fruitful research directions seem to be:
\begin{itemize}
\item generalization of the NS formulas to other root systems;
\item constraining the form of elliptic linear equations from the form of the elliptic
  NS function;
\item understanding the underlying reason why the peculiar appearing combinations of
  $\tgg_{i,j}(k)$ factors are the right terms for direct quantization;
\item combination of key ideas from \cite{paper:MMP-twisted-cher} and the present text
  in order to obtain general formula  for the tBAF.
\end{itemize}

\section*{Acknowledgments}
We are grateful to L. Bishler, N. Tselousov, M. Sharov, M. Chepurnoi, I. Ryzhkov and A. Oreshina
for stimulating discussions.

A.P. gratefully acknowledges support from the Ministry of Science and Higher Education
of the Russian Federation (agreement no.075-03-2025-662).
Our work is also partly supported
by Armenian SCS grants 24WS-1C031 (A.Mir.)
and by the grant of the Foundation for
the Advancement of Theoretical Physics and Mathematics “BASIS".


\begin{thebibliography}{12}

\bibitem{paper:MMP-twisted-cher}
  A. Mironov, A. Morozov, A. Popolitov,
  \texttt{arXiv:2601.10500}

\bibitem{paper:Ch-baf-original}
  O. Chalykh, \textbf{Adv.Math.} 166(2) (2002) 193-259, \texttt{math/0212313}
\bibitem{paper:ES-alg-int}
  P. Etingof, K. Styrkas, \textbf{Compositio Math.} 114 (1998) 125-152, \texttt{q-alg/9603022}

\bibitem{paper:MMP-dim-comm-fams}
  A. Mironov, A. Morozov, A. Popolitov,
  \textbf{JHEP} 2024 (2024) 200, \texttt{arXiv:2406.16688}

\bibitem{paper:DIM}
  J. Ding, K. Iohara, \textbf{Lett. Math. Phys.} 41 (1997) 181-193, \texttt{q-alg/9608002}
  
\bibitem{paper:Miki}
  K. Miki, \textbf{J. Math. Phys.} 48 (2007) 123520

\bibitem{paper:PRS}
  C.N. Pope, L.J. Romans, X. Shen, \textbf{Phys.Lett. } B236 (1989) 173-178;
  \textbf{Nucl.Phys.} B339 (1990) 191-221;
  \textbf{Phys.Lett.} B242 (1990) 401-406; \textbf{Phys.Lett.} B245 (1990) 72-78

\bibitem{paper:FKRW}
  E. Frenkel, V. Kac, A. Radul, W. Wang, \textbf{Comm.Math.Phys.} 170 (1995) 337-358,
  \texttt{hep-th/9405121}

\bibitem{paper:AFMO}
  H. Awata, M. Fukuma, Y. Matsuo, S. Odake, \textbf{Prog.Theor.Phys.Suppl.}
  118 (1995) 343-374, \texttt{hep-th/9408158}

\bibitem{paper:KR-transformation-groups}
  V.G. Kac, A. Radul, {\bf Transformation groups} 1 (1996) 41-70, \texttt{hep-th/9512150}

\bibitem{paper:Ts-yangian}
  A. Tsymbaliuk, \textbf{Adv.Math.} 304 (2017) 583-645, \texttt{arXiv:1404.5240}

\bibitem{paper:P-yangian}
  T. Proch´azka, \textbf{JHEP} 10 (2016) 077, \texttt{arXiv:1512.07178}

\bibitem{paper:MOP-beta-models-directly}
  A. Mironov, A. Oreshina, A. Popolitov,
  \textbf{JETP Lett.} 120 (2024) 62, \texttt{arXiv:2404.18843}

\bibitem{paper:MOP-beta-wlzz}
  A. Mironov, A. Oreshina, A. Popolitov,
  \textbf{EPJC} 84 (2024) 705, \texttt{arXiv:2403.05965}

\bibitem{paper:MMMP-comm-fams}
  A. Mironov, V. Mishnyakov, A. Morozov, A. Popolitov,
  \textbf{JHEP} 09 (2023) 65, \texttt{arXiv:2306.06623}
  
  \bibitem{paper:MMP-basic-triad}
A.~Mironov, A.~Morozov, A.~Popolitov,
{\bf Phys.Lett.} \textbf{B869} (2025) 139840,
arXiv:2411.16517
  
  \bibitem{ChE}   O. Chalykh, P. Etingof, {\bf Advances in Mathematics} 238 (2013) 246-289,
arXiv:1111.0515

\bibitem{Mac} I.G. Macdonald,
  \textit{Symmetric functions and Hall polynomials},
  Oxford University Press, 1995
  
  \bibitem{paper:NS-direct-approach-to-bispectral}
  M. Noumi, J. Shiraishi, \texttt{arXiv:1206.5364}

\bibitem{paper:ChF-baf-twisted}
  O. Chalykh, M. Fairon,
  {\bf J.Geom.Phys.} 121 (2017) 413-437, \texttt{arXiv:1704.05814}

\bibitem{paper:MMP-chalykh-approach-to-dim-eigenfunctions}
  A. Mironov, A. Morozov, A. Popolitov,
  \textbf{Phys.Lett.} B863 (2025) 139380, \texttt{arXiv:2410.10685}

\bibitem{paper:MMPZ-ell-triad}
  A. Mironov, A. Morozov, A. Popolitov, Z. Zakirova,
  {\bf Phys.Lett.} B865 (2025) 139467, \texttt{arXiv:2412.19588}

\bibitem{paper:MMP-chalykh-approach-to-dim}
  A. Mironov, A. Morozov, A. Popolitov,
  {\bf Phys.Lett.} B863 (2025) 139380, \texttt{arXiv:2410.10685}

\bibitem{paper:MMP-baf-as-eigens}
  A. Mironov, A. Morozov, A. Popolitov,
  \textbf{Nucl.Phys.} B1012 (2025) 116809, \texttt{arXiv:2411.14194}

\end{thebibliography}
\end{document}